\documentclass[12pt,preprint]{aastex}
\usepackage{graphicx}

\shorttitle{C, N, and O abundances in Omega Centauri} 
\shortauthors{A.\ F.\ Marino} 
\usepackage{ulem}

\def\etal{\mbox{\rm et al.\ }}

\def\spro{\mbox{$s$-process}}

\def\teff{\mbox{T$_{\rm eff}$}}
\def\logg{\mbox{log~{\it g}}}
\def\vmicro{\mbox{$\xi_{\rm t}$}}

\begin{document}

\title{
  The C+N+O abundance of $\omega$~Centauri giant stars:
  implications on the chemical enrichment scenario and the relative ages of different stellar populations
          \thanks{Based on data collected at the European Southern
    Observatory with the FLAMES/GIRAFFE spectrograph.           }}

\author{
A.\ F. \,Marino\altaffilmark{1}, 
A.\ P. \,Milone\altaffilmark{2,3}, 
G.\ Piotto\altaffilmark{4}, 
S.\ Cassisi\altaffilmark{5}, 
F.\ D'Antona\altaffilmark{6}, 
J.\ Anderson\altaffilmark{7}, 
A.\ Aparicio\altaffilmark{2,3}, 
L.\ R. \, Bedin\altaffilmark{8}, 
A.\ Renzini\altaffilmark{8},
S.\ Villanova\altaffilmark{9}
}

\altaffiltext{1}{Max-Planck-Institut f\"{u}r Astrophysik, Karl-Schwarzschild-Str. 1,
             85741 Garching bei M\"{u}nchen, Germany\\
             \email{amarino@MPA-Garching.MPG.DE}}

\altaffiltext{2}{Instituto de Astrof\'isica de Canarias, E-38200 La
               Laguna, Tenerife, Canary Islands, Spain\\
               \email{milone@iac.es, aparicio@iac.es}}
 
\altaffiltext{3}{Departamento de Astrof\'isica, Universidad de La Laguna, 
           E-38200 La Laguna, Tenerife, Canary Islands, Spain}

\altaffiltext{4}{Dipartimento  di   Astronomia,  Universit\`a  di Padova,
             Vicolo dell'Osservatorio 3, Padova, I-35122, Italy\\
             \email{giampaolo.piotto@unipd.it}}

\altaffiltext{5}{INAF-Osservatorio Astronomico di Teramo, Via
  M. Maggini, 64100 Teramo, Italy\\
\email{cassisi@oa-teramo.inaf.it}}

\altaffiltext{6}{INAF-Osservatorio Astronomico di Roma, via Frascati 33, I-00040 Monteporzio, Italy\\
\email{dantona@oa-roma.inaf.it}}

\altaffiltext{7}{Space Telescope Science Institute, 3700 San Martin
 Drive, Baltimore, MD 21218, USA\\
\email{jayander@stsci.edu}}

\altaffiltext{8}{INAF-Osservatorio Astronomico di Padova, Vicolo
 dell’Osservatorio 5, 35122 Padova, Italy\\
\email{luigi.bedin@oapd.inaf.it, alvio.renzini@oapd.inaf.it}}

\altaffiltext{9}{Departamento de Astronom\'ia, Universidad de Concepci\'on, Casilla 160-C, Concepci\'on, Chile\\
\email{svillanova@astro-udec.cl}}

\begin{abstract}
We present a chemical-composition analysis of 77 red-giant stars 
in Omega Centauri.  We have measured abundances for carbon and 
nitrogen, and combined our results with abundances of O, Na, La, and Fe 
that we determined in our previous work.  Our aim is to 
better understand the peculiar chemical-enrichment history 
of this cluster, 
by studying how the total C+N+O content varies among 
the different-metallicity stellar groups,
and among stars at different places along the
                   Na-O anticorrelation.
We find the (anti)correlations among the light
                     elements that would be expected 
on theoretical ground for matter that has
been nuclearly processed via high-temperature proton captures.
The overall [(C+N+O)/Fe] increases by $\sim$0.5 dex from [Fe/H]$\sim -$2.0 
to [Fe/H]$\sim -$0.9.  
Our results provide insight into the 
  chemical-enrichment history of the cluster, and the measured 
  CNO variations provide important corrections for estimating 
  the relative ages of the different stellar populations.
\end{abstract}


\section{Introduction}
\label{introduction}
Omega~Centauri ($\omega$~Cen) is one of the most intriguing
Globular Clusters (GCs) of the Galaxy.  At odds with the majority of GCs, 
which are mono-metallic, it shows large star-to-star metallicity variations 
up to more than one dex (e.g.\ Norris, Freeman \& Mighell 1996, 
Suntzeff \& Kraft 1996).  
At the same time, it shares with the most mono-metallic 
GCs the presence of a Na-O anticorrelation, 
which is present across almost the entire metallicity range
(Marino \etal 2011a, 
Johnson \& Pilachowski 2010).
These two facts suggest unique complexity in the cluster's 
      star-formation history, and its chemical evolution.

The complexity of $\omega$~Cen also manifests itself in its color-magnitude 
diagram (CMD) with the presence of multiple red-giant (RGBs), multiple 
sub-giant branches (SGBs, Lee \etal 1999, Pancino \etal 2000), and an 
extended horizontal branch (Villanova \etal 2007; Cassisi \etal 2009, 
D'Antona, Caloi \& Venura 2010; Bellini \etal 2010).  The main sequence (MS) is double (Anderson 1997), 
and a third, less-populated MS has been discovered by Bedin \etal (2004) 
and associated with the most metal-rich population.  The bluer MS (bMS) has a higher 
metallicity than the red MS (rMS, Piotto \etal 2005).  So far, the only way 
to account for these observations is to assume the bMS to be highly 
enhanced in He (Norris 2004; Bedin \etal 2004; Piotto \etal 2005).
Direct evidence in this directions
has been recently
  detected from the analysis of the He I 10830 transition on
  $\omega$~Cen RGB stars (Dupree \etal 2011). 
  The correlation of the He line detection
  with [Fe/H], Al and Na supports the assumption that helium is
  enhanced in the bMS.

Due to the observational scenario, more complex than in any other GC,
it has often been suggested
(e.g.\ Bekki \& Freeman 2003) that $\omega$~Cen may be
the remnant of a now-dissolved dwarf 
galaxy, once similar of the Sagittarius 
dwarf (with its central globular cluster M54) now being thorn apart by the 
Galactic tidal field.
In any case, a successful description of the $\omega$~Cen star-formation history 
should be able to explain both the Na-O anticorrelation at the various
metallicities
 and the general rise 
          in slow-process (\spro) elements as a function of metallicity
(Norris \& Da Costa 1995; Johnson \& Pilachowski 2010; 
Smith \etal 2000; Marino \etal 2011a).  
The presence of the Na-O anticorrelation implies that $\omega$~Cen, similarly to 
mono-metallic GCs, has experienced enrichment from high-temperature 
H-burning processed material.  At the same time, an additional physical 
mechanism should be present to produce \spro\ elements.
In the Sun, the \spro\ elements abundance is mainly due to
  two components: the main component (attributed to low-mass AGB stars
  of 1.5-3${\rm M_{\odot}}$; Busso, Gallino \& Wasserburg 1999) and the weak component (attributed to massive stars; see Raiteri et al. 1993, and references therein). At solar metallicity, this last component mainly produces the lighter \spro\ nuclei, at most up to Sr and Kr (Raiteri et al. 1993).
In the case of $\omega$~Cen, if we assume that 
the $s$-elements are produced in the strong component, i.e. in 
a population of less massive
asymptotic-giant-branch (AGB) stars (Busso \etal 1999, Ventura \etal
2009), we have to face a timescale discrepancy that makes hard to solve the entire puzzle (D'Antona \etal 2011).
Indeed, these AGB have longer lifetimes than the more massive AGB stars, 
which are presumed responsible for the He enrichment and for Na-O 
anticorrelations.
If, alternatively, the \spro\ elements are produced in the weak component
by massive stars that explode as Type II Supernovae,  
it remains to be seen whether the models can produce elements as heavy
as Ba and La in stellar environments characterized by the chemical
abundances observed in $\omega$~Cen. 

An important ingredient to understand the star-formation history of this 
poorly-understood GC is the determination of the relative ages of its 
stellar populations.  Numerous studies based on CMD analysis combined 
with metallicity distribution of turnoff (TO) and SGB stars have yielded 
conflicting results, suggesting age differences from less than 2~Gyr 
(Lee \etal 2005, Sollima \etal 2005, Calamida \etal 2009),  
up to 5~Gyrs 
(Villanova \etal 2007).  
Rapid-formation scenarios wherein the entire cluster could have formed 
within a few times $10^{8}$ years have recently been suggested 
by D'Antona \etal (2011) and Valcarce \&
Catelan (2011).  

Theoretical isochrones show that a variation 
      of the total C+N+O abundance can have an 
effect 
      on the GC ages obtained from CMD fitting.
Therefore, the 
relative age-dating of the stellar populations hosted in the cluster would 
be severely affected by the presence of C+N+O differences among the 
$\omega$~Cen  sub-populations (Cassisi \etal 2008, Ventura \etal
2009, Pietrinferni \etal 2009, D'Antona \etal 2009). 

In an effort to shed light on these issues, in this paper we measure the 
overall C+N+O abundance in 77 RGB stars of $\omega$~Cen, which span nearly 
its entire metallicity range.  
The layout of this paper is as follows:
in Section~\ref{sec:data} we describe the data
analysis; results are presented in Section~\ref{sec:results}, and their impact 
on the chemical enrichment scenario and the age measurements are
discussed in Section~\ref{sec:scenarios} and Section~\ref{sec:discussion},
respectively; Section~\ref{sec:conclusions} is a summary of our results.

\section{Observations and data reduction}
\label{sec:data}

Our data-set consists in a sample of 77 RGB stars observed with the
FLAMES/GIRAFFE HR4 setup (program: 082.D-0424A).  We also have, from Marino \etal (2011a), 
additional spectra for the same stars with different GIRAFFE setups, from which
we have derived abundances for Fe, Na, O, and $n$-capture element La
(75\% \spro\ in the solar system, Simmerer \etal 2004).
In Marino \etal (2011a) we provide meaurements also for Ba
abundances, but 
here we prefer to use only La as representative of  $n$-capture elements
because Ba measurements are more uncertain, since the only analyzed Ba
transition is a blend (see Marino \etal 2011a).
We refer the reader to this paper for a more detailed description of the sample 
and the data reduction.  Here we analyse, for 77 stars of the Marino \etal (2011a) sample, 
chemical abundances for carbon and nitrogen.  

The HR4 setup covers the spectral range from $\sim$4188 to $\sim$4392~\AA, providing 
a resolution $R \sim$20,000.  The typical signal-to-noise ratio of the final 
combined spectra at the central wavelength of the spectral range is $\sim$80.

\begin{figure*}[ht!]
\centering
\epsscale{.65}
\plotone{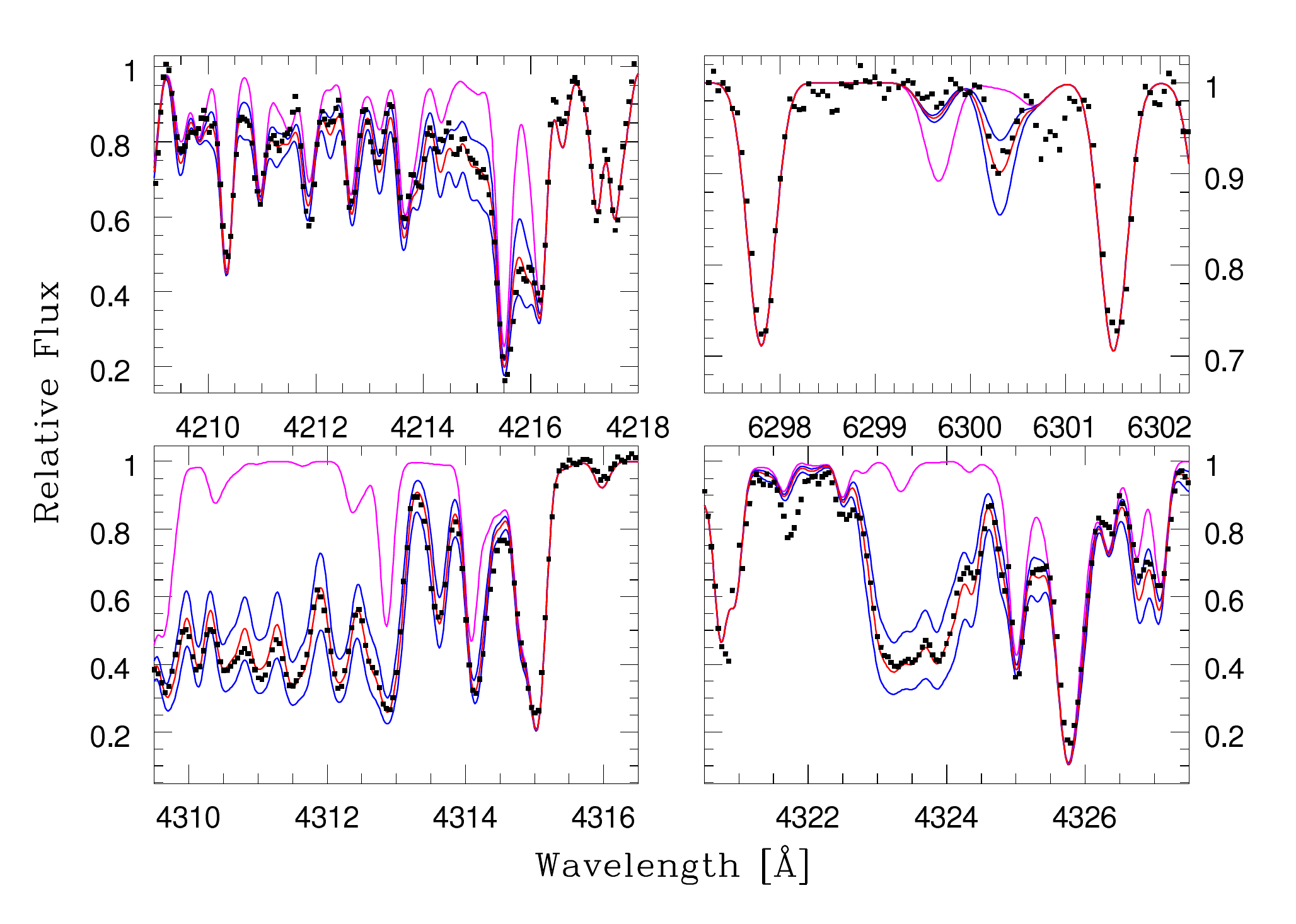}
\caption{Observed and synthetic spectra around CH-bandheads (lower panels), 
         CN-band (upper-left panel) and O line (upper-right panel) 
         for the star \#248664.  The magenta line is the spectrum computed with 
         no contribution from C, N, and O; the red line and blue lines
         are the best-fitting and 
         the syntheses computed with 
         abundance altered by $\pm$0.25 dex from the best value, respectively.}
\label{sintesi}
\end{figure*}

Our abundance analysis used the local thermodynamic equilibrium analysis code 
MOOG (Sneden 1973).  Carbon was measured from spectral synthesis of the G-bandheads
($A^2\Delta-X^2\Pi$) near 4314 and 4323~\AA.  The nitrogen abundance was derived 
from synthesis of the CN blue-system ($B^2\Sigma-X^2\Sigma$) bandhead at 
$\sim$4215~\AA.  The synthesis line list for the blue CN band is described in 
Hill \etal (2002).  The line list for the CH band was provided by B. Plez 
(private communication).
Synthetic spectra were employed 
by using Castelli \& Kurucz (2004) models for effective temperatures (\teff), 
gravities (\logg), metallicities, and microturbolences (\vmicro) determined in 
Marino \etal (2011a).  In computing the C abundance we used the previously determined O contents 
(Marino \etal 2011a), and for N, both observed C and O abundances needed to be employed. 
As an example, we show in Figure~\ref{sintesi} the spectral synthesis for the 
CH and CN bands, and the O line for star \#248664.  For more details on 
the atmospheric-parameter determinations, and a discussion on related 
uncertainties, we refer to Marino \etal (2011a).  Here, we focus on the C and N measurements.

The error analysis for C and N was performed by varying \teff, \logg, 
metallicity, and \vmicro\ for several stars, and redetermining the abundances. 
The parameters were varied by $\Delta$\teff=$\pm$100 K, $\Delta$\logg=$\pm$0.20, 
$\Delta$\vmicro=$\pm$0.10~km/sec, and metallicity by 0.10 dex.  In addition, 
the derived C abundance is dependent on the O abundance and therefore so is 
the N abundance, and the circulation of correlated errors tends to mimic 
the effects of CNO-cycle burning (e.g.\ Iben \&
Renzini 1983).  
Indeed, in the red giant atmospheres carbon is mainly present in the
CO molecules, and in the molecular equilibrium an over-estimation in
oxygen produces an over-estimation of carbon from the CH band, and viceversa.
In the CN bands, a carbon over-estimation causes nitrogen to be under-estimated.

In the treatment 
of the internal errors for carbon, we then varied the 
oxygen abundance by 
$\pm$0.10, and repeated the spectrum synthesis to determine the error 
introduced by uncertainties in the O abundances.  These errors were then 
added in quadrature to the errors introduced by 
atmospheric parameters 
and yielded an overall error of $\sim \pm$0.10 dex to the [C/Fe] abundances.
For nitrogen we varied both O and C, the latter by $\pm$0.10 dex.  We added 
these errors in quadrature with those introduced by the model atmosphere and
estimated the internal uncertainty of the [N/Fe] values to be $\sim$0.20-0.30.

Carbon and nitrogen abundances for $\omega$~Cen RGB stars were determined by 
Brown \& Wallerstein (1993; 6 stars), Norris \& Da Costa (1995, 40 stars), 
and Stanford, Da Costa, \& Norris (2010; 33 stars).  A proper comparison 
with these studies cannot be made because there are no stars in common,
however we note that our C and N values span a range similar to that of 
Brown \& Wallerstein (1993) and Stanford \etal (2010).  Possible 
systematic differences in C could not be excluded between our stars and 
those of Stanford et al., while systematically lower N abundances 
(of $\sim$0.50 dex) were found by Norris \& Da Costa (1995).

In Marino \etal (2011a) we derived the O abundance from the line at
6300 \AA\ by comparing our observations with synthetic 
spectra that had solar C and N.  Here, we
redetermine oxygen 
by adopting in the synthesis the measured N and C abundances.  
These improved [O/Fe] measures are in good agreement with the previous 
measurements, with only few stars differing by more than 0.1 dex from the 
values in Marino \etal (2011a). 

%

\section{Results}
\label{sec:results}

\begin{figure*}[ht!]
\centering
\epsscale{.8}
\plotone{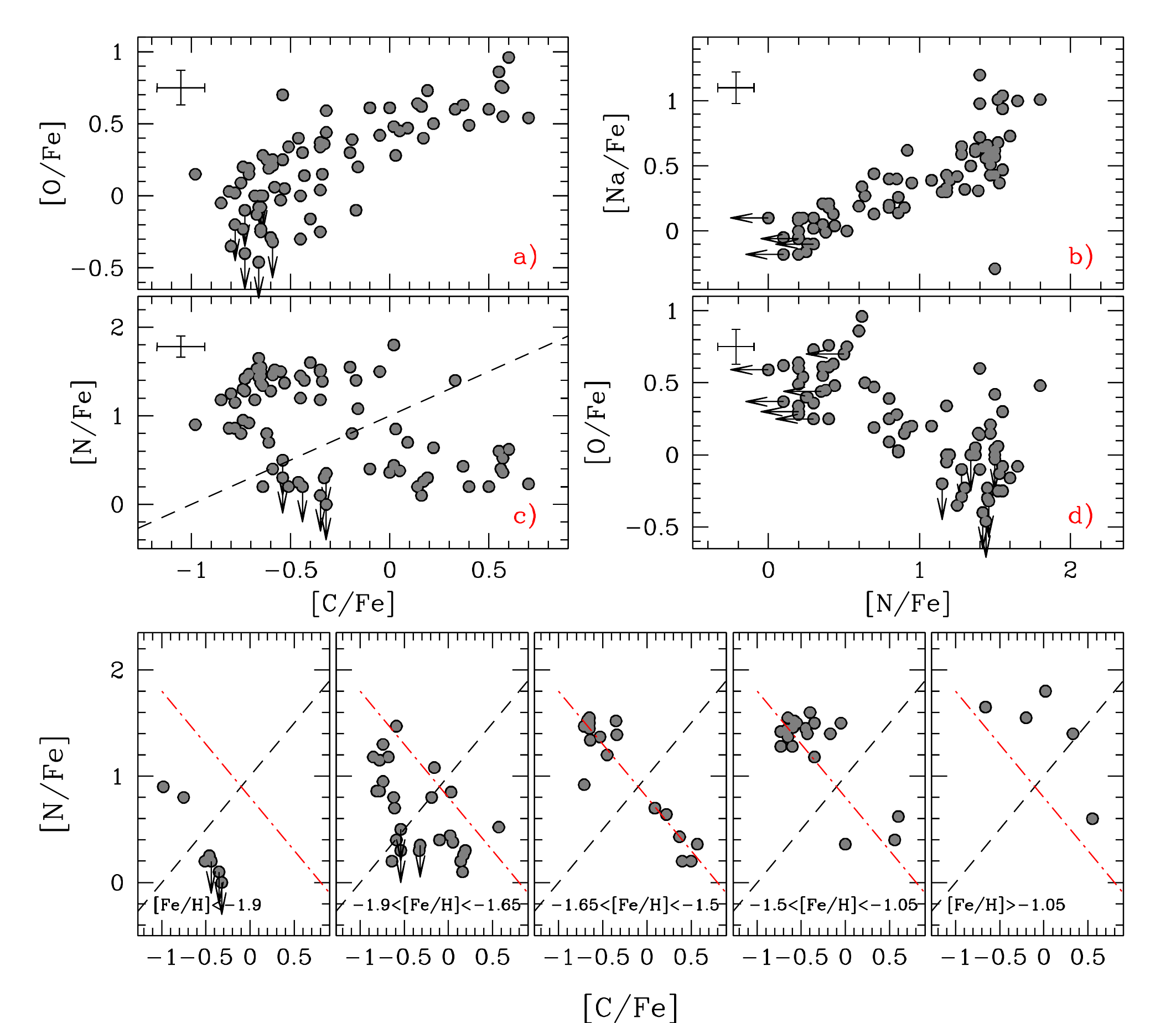}
\caption{
   \textit{Upper panels}: 
        [N/Fe] and [O/Fe] vs.\ [C/Fe]  (panels c, a), 
        and [O/Fe] and [Na/Fe] vs.\ [N/Fe] (panels d, b).
        The dashed black line in panel (c) separates C-poor/N-rich stars from 
        C-rich/N-poor.  
   \textit{Lower panels}: [N/Fe] vs.\ [C/Fe] in metallicity bins
        defined in Marino \etal (2011a).  In each panel, the dashed black line is 
        the same as in panel (c), the red dot-dashed line represents 
       the "best-fit" line of constant [C/Fe]+[N/Fe] (=0.80) for the mid-metallicity stars.
} 
\label{ele}
\end{figure*}
%
Figure~\ref{ele} plots the abundances of C, N, and O
(Tab.~1)
measured in this paper 
and the Na measurements from Marino \etal (2011a).  We observe large star-to-star variations 
in all of these elements, as has already observed in all the GCs studied 
to date.  Carbon spans a range from [C/Fe]$\sim -$0.9 to [C/Fe]$\sim +$0.6.
Most stars are C-depleted, but there is a population of stars with [C/Fe]$>$0.0.
Nitrogen spans almost 2 dex, from [N/Fe]$\sim$0.0 
to stars with [N/Fe]$\sim$1.8 dex.  
There is only one star (\#214842) that 
                  is enhanced in both C and N (Figure~\ref{ele}, panel c).

Our results 
show the existence of well-defined patterns among C, N, O,
and Na.  A positive correlation is present between O and C, and between 
Na and N, and N anticorrelates with O (Figure~\ref{ele}, panels a, b, d).
A C-N anticorrelation is not obvious (panel c), but it can be more easily 
recognizable by grouping stars on the basis of different Fe content,
as will be discussed in Section~\ref{cnofe}.
Such chemical patterns indicate the occurrence of high-temperature
       H-burning involving the CNO and NeNa cycles, and are consistent with
       prediction
in Ventura \& D'Antona (2009). 

In the following sections we present the pattern of C, N, O
  abundances with [Fe/H] (Section~\ref{cnofe}) and with respect to the
  position of stars on the Na-O anticorrelation (Section~\ref{cnoona}) studied by Marino
  \etal (2011a).

\begin{table}[ht!]
\label{abundances}
\centering
\begin{tabular}{lrrr}
\hline 
\hline
ID         &   [C/Fe] & [N/Fe] & [O/Fe]\\
\hline
246266 &    $-$0.68 &1.18&   0.00\\ 
251010 &    $-$0.62 &0.80 &  0.25\\
249334 &   $-$0.80  &1.25& $-$0.35\\ 
248664 &    $-$0.19 &0.80 &   0.39 \\
240777 &      0.02     &0.44 &   0.48 \\
239549 &    $-$0.71 &0.92 &  0.19\\
\hline
\end{tabular}
\caption{Carbon, nitrogen and oxygen abundances ratios relative to
  iron for the analyzed stars The full table is only electronically available.}
\end{table}

\subsection{CNO pattern with [Fe/H]}
\label{cnofe}
Before discussing in detail the C, N, O patterns, we recall that $\omega$~Cen 
exhibits a large spread in [Fe/H], ranging from $\sim-2.0$ to $\sim -0.7$ dex.  
The iron distribution is multi-modal with distinct peaks at [Fe/H]$\sim -$1.75, $-$1.60, 
$-$1.45, $-$1.00, 
and there is a broad 
       distribution of stars extending up to
[Fe/H]$\sim -$0.70 (Marino \etal 2011a).  
Although the existence of these multiple populations is clear,
observational errors often prevent us from definitively 
establishing which individual stars belong to which population.
Keeping this in mind, Marino \etal (2011a) defined 
six groups of stars with different iron abundances to investigate the Na-O anticorrelation 
for different metallicity groups (see their Figure~8). 

In the lower panels
of Figure~\ref{ele}, we perform a similar analysis for C versus N in the same 
metallicity bins defined in Marino \etal (2011a).  
Unfortunately, we do not have C and N
      measurements for the most metal-rich
group, with [Fe/H]$>-$0.90, of Marino \etal (2011a).  
We see that the C-N anticorrelation is present 
across the entire range of metallicities.  
The dashed-black line in each panel (that is the same as in panel
  c) has been traced to arbitrarily separate the C-poor/N-rich
  population from the C-rich/N-poor one. 
The red dot-dashed line is the ''best-fit'' line of constant [C/Fe]+[N/Fe]
for the mid-metallicity stars, that has been taken equal to 0.80.

It is worth noting that when we 
compare the relative location along the C-N anticorrelation with 
this line, 
it becomes evident that the average 
C abundance increases from lower to higher metallicities. 
The shape of the 
C-N anticorrelation for different metallicities is similar to that of 
the Na-O anticorrelation (Figure~8 of Marino \etal 2011a), with the most metal-rich groups 
hosting a larger fraction of N-rich (Na-rich, O-C-poor) stars.
This behavior leads few stars, belonging to the metal richer
  group (e.g.\ \#214842), to be enhanced both in C and N.

\begin{figure*}[ht!]
\centering
\epsscale{.75}
\plotone{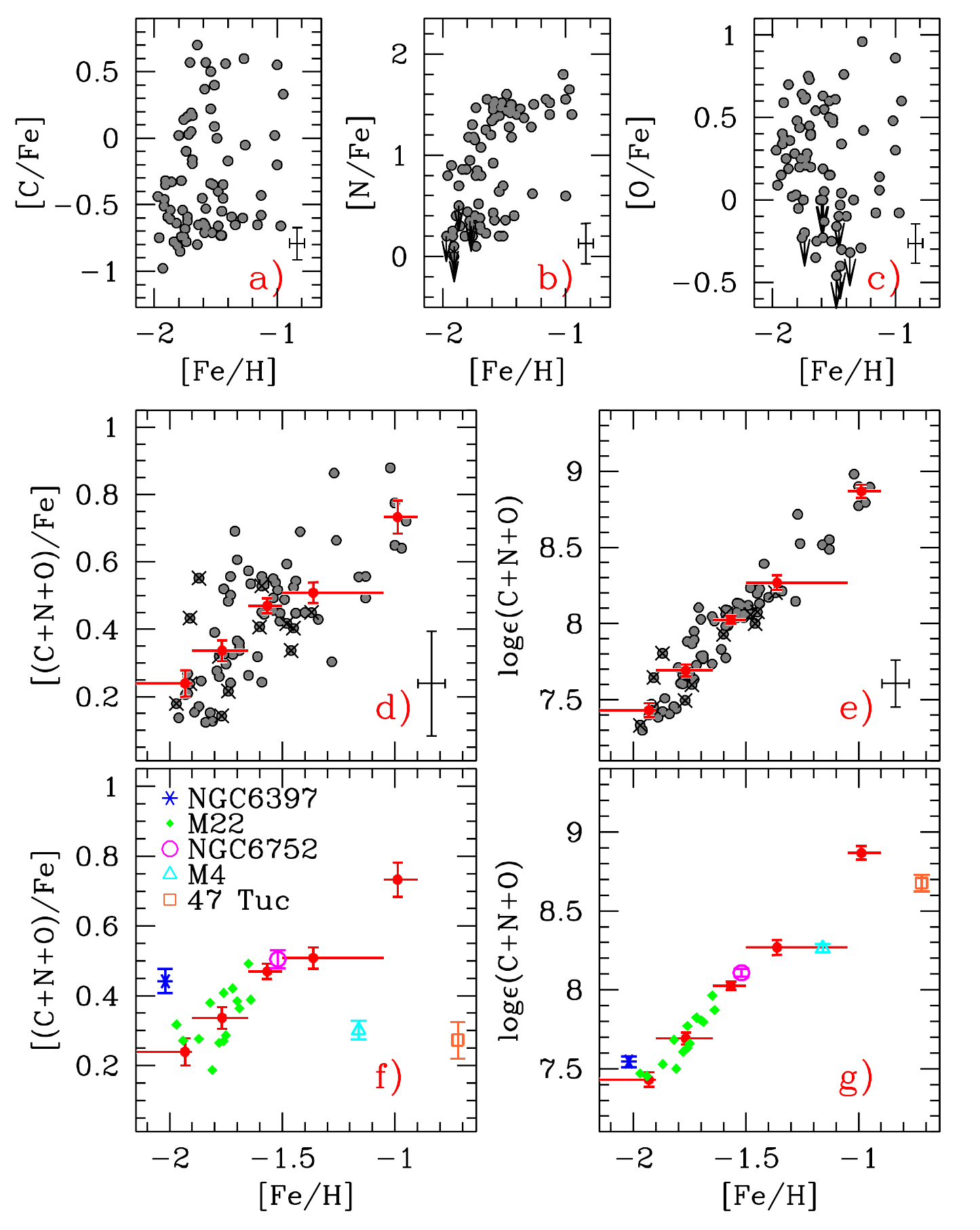}
\caption{
   \textit{Upper panels}: 
      [C/Fe], [N/Fe], and [O/Fe] vs.\ [Fe/H]. 
   \textit{Middle panels}: 
      [(C+N+O)/Fe] and log$\epsilon$(C+N+O) vs.\ [Fe/H].  Stars with 
     upper limits for O or N abundances are represented as grey crosses.
      Red points represent the average C+N+O content in the metallicity
      intervals spanned by the horizontal red bars. Vertical bars are the 
      errors associated with the mean values.
   \textit{Lower panels}: Average [(C+N+O)/Fe] vs.\ iron for the GCs 
      quoted in the inset compared with the $\omega$~Cen CNO mean values 
      (red points) of panels $d$ and $e$. Data for the
      mono-metallicity GCs are from Ivans \etal (1999) for M4, and
      Carretta \etal (2005) for  47~Tuc, NGC~6752, NGC~6397. Data for
    M22 are from Marino \etal (2011b).} 
\label{elvsFe}
\end{figure*}
%

   In the upper panels of Figure~\ref{elvsFe}, we plot the individual carbon, 
   nitrogen, and oxygen abundances against iron.  While [N/Fe] 
   exhibits a mild correlation with [Fe/H], there is no clear correlation 
   either between [C/Fe] and [Fe/H] or between [O/Fe] and [Fe/H].
A significant spread in light elements is present at all metallicities
with [C/Fe], [N/Fe], and [O/Fe] spanning large ranges (more than 1 dex).

   If, however, we instead plot the {\it total} [C+N+O] against [Fe/H], 
   we find a correlation with a Pearson coefficient of 0.71. 
  This is shown in panel (d) 
   for the abundance relative to iron and in panel (e) for the total 
   abundance.  
The red points show the 
   averages over bins in [Fe/H] defined above.  The vertical error bars correspond
   to the error in the mean related to the observed dispersion and 
   the horizontal bars denote the boundaries of the [Fe/H] bins.  

   The representative error bar shown in the lower right of panels (d) 
   and (e) comes from adding the individual C, N, and O errors in 
   quadrature, and indicates that the error in the total [(C+N+O)/Fe]
   should be quite large ($\sim$0.3 dex).  When we estimated the errors in 
   these abundances, we made the assumption that the abundances were 
   statistically independent.  Given, though, that the 
   C, N, and O abundances are correlated with each other, a more realistic 
   estimate of the error might be derived empirically from the rms of 
   the [(C+N+O)/Fe] within each metallicity bin.  In this case, we find 
   a rms ranging from 0.10 to 0.15 dex, and stars with
intermediate metallicity ($-$1.65$\leq$[Fe/H]$< -$1.05) have the largest 
rms.

   The bottom panels of Figure~\ref{elvsFe} show the trends we have derived for
   $\omega$~Cen in the context of other clusters.  M22 has stars with
   a range of metallicities (Marino \etal 2009), and these stars are shown as the green
   points, which happen to follow the $\omega$~Cen trend almost perfectly.
   On the other hand, the mono-metallic clusters for which we have 
   data (NGC~6397, NGC~6752, M4, and 47~Tuc) do not follow this trend 
   and, in general, show no discernible trend of [(C+N+O)/Fe] with 
   metallicity.

\subsection{CNO, Na and La abundances in the Na-O anticorrelation}
\label{cnoona}
To investigate how the studied chemical abundances 
behave for stars occupying a different location on the Na-O plane, 
we divided stars into two groups based on their location 
   along the Na-O trend (Marino \etal 2011a). 
  In normal mono-metallic clusters different generations of stars can
   be segregated in this way, and
  the Na-O groupings tend to
   have different C and N abundances, as can be seen in the case of 
   M4 (Marino et al 2008, see their Fig 10).  The situation in
   $\omega$~Cen has an added level of complexity, since the large variation in
   metallicity does not allow us to simply identify two (or more) populations via
   Na and O.
Both Johnson \& Pilachowski (2010) and Marino \etal
   (2011a) have shown that stars of both {\it first}
   and {\it second}
 generations are present across a large range of
   metallicities.  
Note that here we extend the same nomenclature used for mono-metallic GCs to
 $\omega$~Cen and
 name first and second generation Na-poor/O-rich and Na-rich/O-poor
 stars respectively, whatever is their [Fe/H]. Of course, in the case of $\omega$~Cen this
 designation should not be taken literally, as stars of both groups
 are present in a large range in metallicity.
We will therefore first follow the chemical patterns
for the elements affected by $p$-captures (C, N, O, Na), $n$-capture elements (La),
and C+N+O  
  for the first generation alone, and then 
   will examine the second generation.

\begin{figure*}[ht!]
\centering
\epsscale{1.0}
\plotone{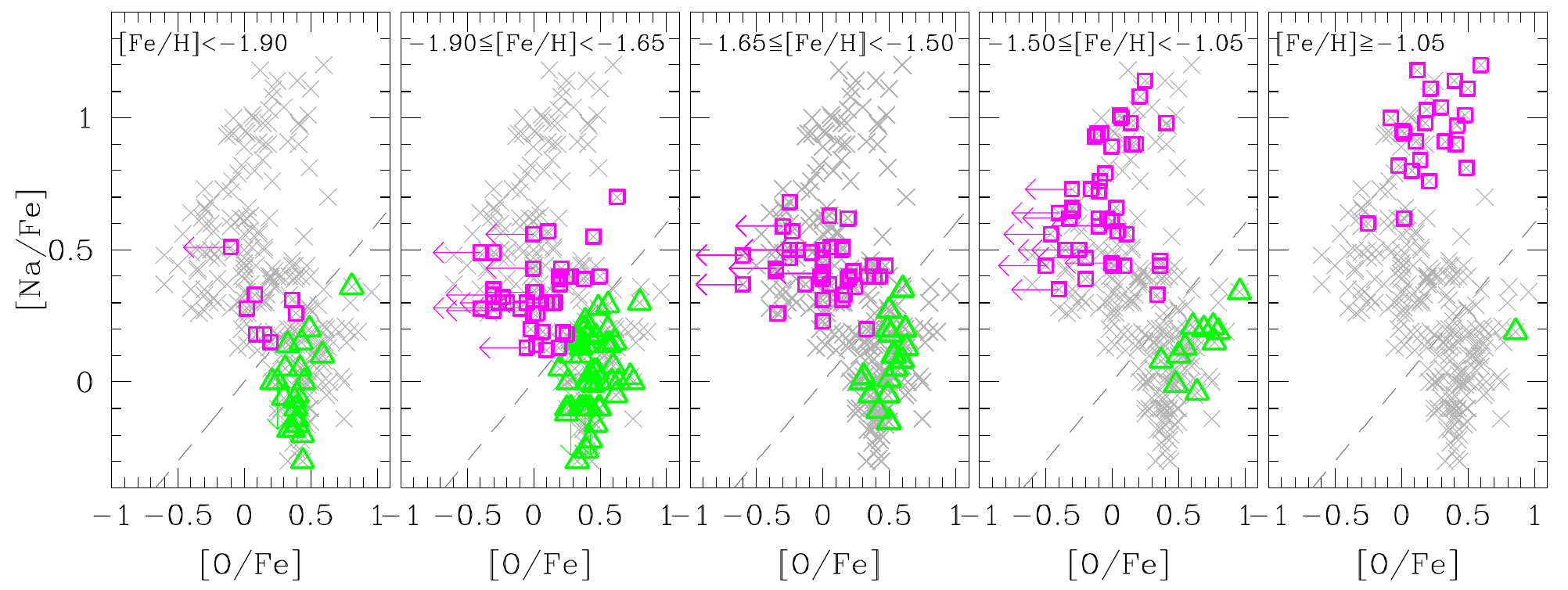}
\caption{Our adopted division of stars in
O-rich/Na-poor ({\it first generation}) and O-poor/Na-rich ({\it
second generation}) has been represented in
the Na-O plane for each metallicity bin,
as quoted in the insets. The dashed line separates the selected
first generation stars represented by green triangles, from the
selected second generation stars represented by magenta squares. Gray
crosses in each panel represent the entire sample analysed by Marino
\etal (2011a). }
\label{ola}
\end{figure*}

\begin{figure*}[ht!]
\centering
\epsscale{0.97}
\plotone{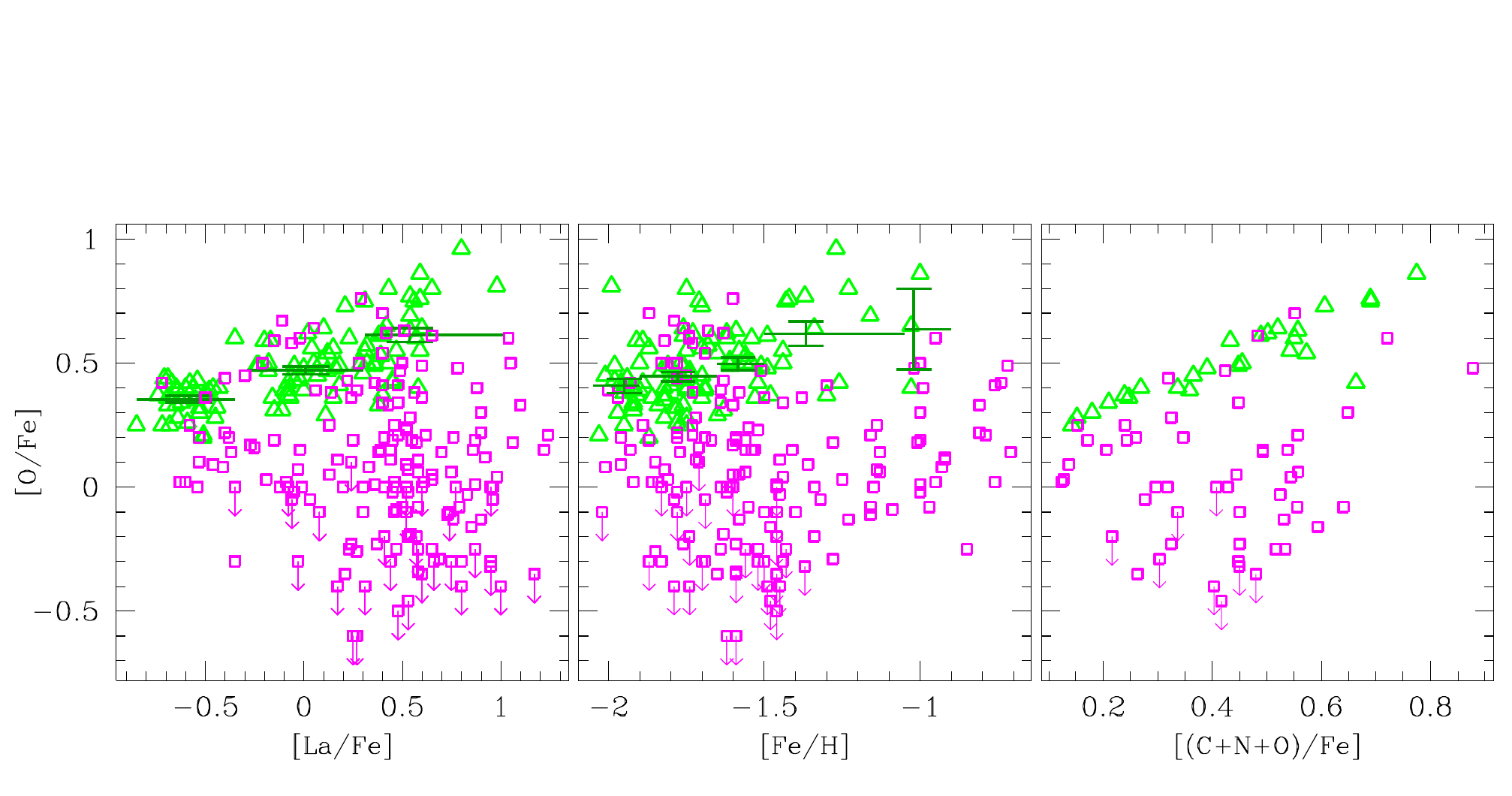}
\caption{O abundance ratios as a function of [La/Fe], [Fe/H] and
 [(C+N+O)/Fe]. Symbols are as in Figure~\ref{ola}.
 The dark green vertical error bars represent the error associated with
the mean [O/Fe] abundance in different intervals in [La/Fe] and [Fe/H] delimited by the
horizontal line.}
\label{ofecno}
\end{figure*}

In Figure~\ref{ola} we represent in different colors
our selected first and second generation stars on the Na-O
anticorrelation from Marino \etal (2011a).
The Na-poor/O-rich first generation (green open triangles) and the
Na-rich/O-poor second generation stars (magenta open squares), 
have been
arbitrarily selected by the dashed line. 
Similarly to what done in Figure~2, stars have been represented in different panels
depending on their [Fe/H] bin. 
In each panel, the grey crosses represent the entire sample.
The so selected first and second generation stars have been plotted in a
O-La plane, as shown in Figure~\ref{ofecno} (right panel).
It is worth noting that in first generation stars 
[O/Fe] abundances correlate with [La/Fe], with  
the Pearson coefficient equal to 0.73.
To quantify this correlation
we have determined the mean [O/Fe]
abundances for three La intervals spanned by first generation stars, whose sizes are indicated by the
horizontal dark-green lines in Figure~\ref{ofecno}.
The mean [O/Fe] values for the different Fe and La intervals are listed in Tab.~2,
togheter with the rms and associated errors.  
In first stellar generation, some hints for a correlation may be present also among [O/Fe] and [Fe/H],
as shown in the middle panel of Figure~\ref{ofecno}. 
In this case the O rise is 
more difficult to be claimed over observational errors 
and the mean [O/Fe] values in the selected metallicity bins
(shown in  dark-green in the middle panel of Figure~\ref{ofecno}) agree
within a 3$\sigma$ values.

\begin{table*}[ht!]
\scriptsize
\label{tab00}
\centering
\begin{tabular}{lcccccc|ccc}
\hline 
\hline
                       & \multicolumn{6}{c}{[Fe/H] range}                 &     \multicolumn{3}{c}{[La/Fe] range}                \\

                       &$<-1.90$ & $-1.90/-1.65$  &$-1.65/-1.50$&$-1.50/-1.05$&$-1.05/-0.95$&$>-0.95$&$<-0.30$&$-0.30/0.30$&$>0.30$   \\\hline
${\rm [O/Fe]}$& 0.41         &0.45                    &0.50                 &0.62                &0.64                &\nodata   &0.35&0.47&0.61    \\
$\pm$            &0.03          &0.02                    &0.03                 &0.05                &0.16                &\nodata   &0.01&0.02&0.03   \\
$\sigma$       &0.13           &0.13                    &0.10                 &0.18                &0.23                &\nodata   &0.08&0.10&0.15    \\
\#~I-gen.~stars  &20&48                 &17                   &14                   &3                      &0             &33&38&31    \\
\hline
\end{tabular}
\caption{Oxygen mean content for first generation stars in the
 Fe bins  of Marino \etal (2011a), and for three intervals in [La/Fe].}
\end{table*}

We suggest some caution with the O trend in first generation stars, as
we cannot fully exclude that it may be due to some unidentified
systematics with iron. Johnson \& Pilachowski (2010) do not find a
trend of [O/Fe] with [Fe/H], but claim instead to find weak trends
($\sim 0.1$ dex) in other $\alpha$ element ratios, such as [Ca/Fe]
and [Si/Fe], which however are even weaker than our trend in
[O/Fe]. Notice that such trend is detectable only among first
generation stars, as in second generation stars oxygen has been
depleted by $p$-capture reactions.

\begin{figure*}[ht!]
\centering
\epsscale{0.7}
\plotone{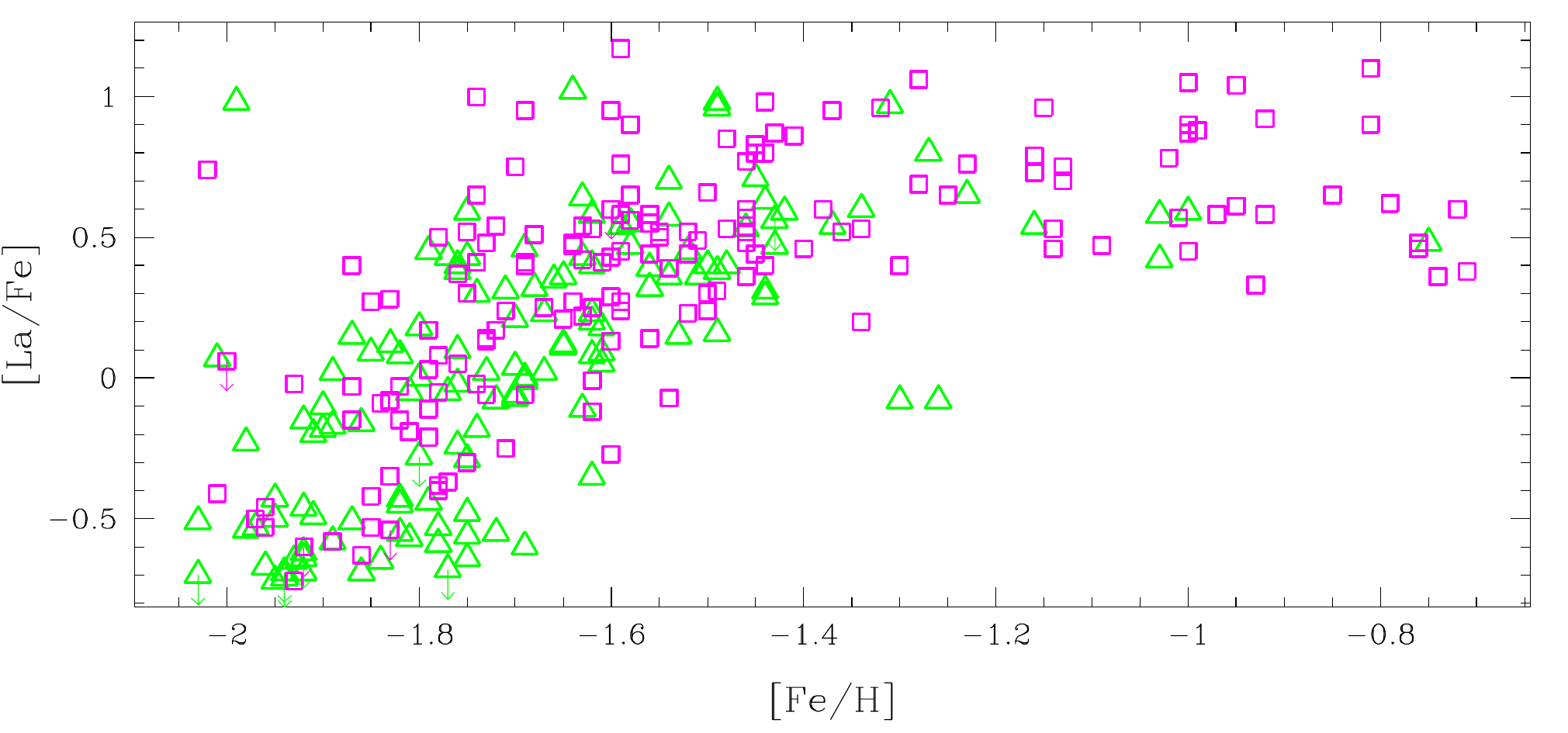}
\caption{[La/Fe] as a function of [Fe/H]. Symbols are as in Figure~\ref{ola}.}
\label{lao}
\end{figure*}

As shown in the right panel of Figure~\ref{ofecno},  in first generation
stars O increases in concert with the total CNO. 
At odds with first generation stars, the second generation ones
(represented in magenta) do not appear to show any correlation with
either La, Fe, and CNO, and occupy a spread region in all these
abundance planes.
Lanthanum abundance ratios follow a similar pattern for first and
second generation stars as shown if Figure~\ref{lao}.
This implies that the material out of which second generation stars
have formed was not exposed to neutron sources (no additional
$n$-captures besides those experienced by first generation stars) but
only to $p$-captures. If AGB stars were responsible for the
$p$-capture processing, this must have taken place in stars
experiencing negligible third dredge-up, otherwise the material would
have been further enriched in \spro\ elements.

\begin{figure*}[ht!]
\centering
\epsscale{0.97}
\plotone{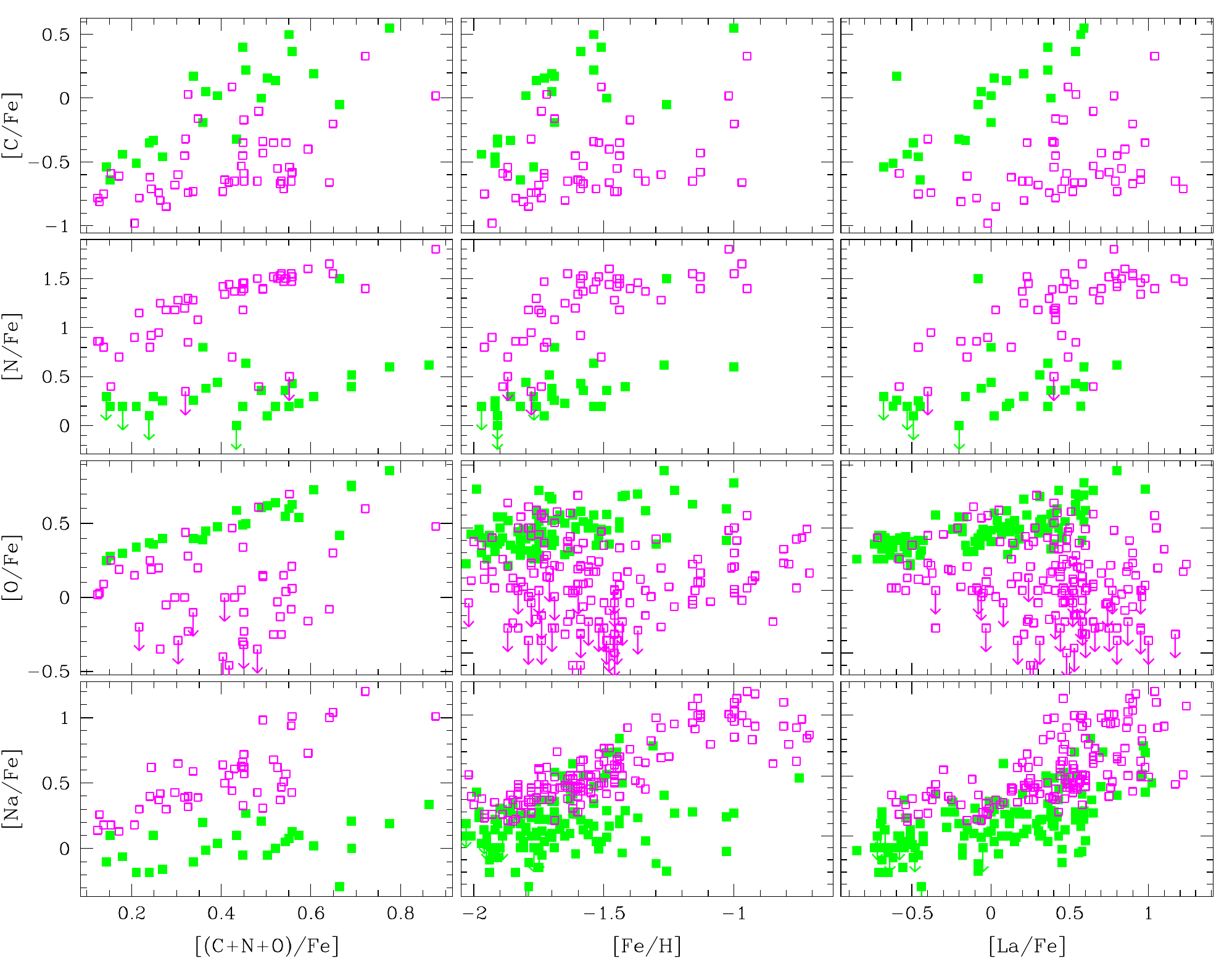}
\caption{C, N, O, and Na abundance ratios as a function of
 [(C+N+O)/Fe], [Fe/H] and [La/Fe]. Symbols are as in Figure~\ref{ola}. }
\label{ona_cno}
\end{figure*}

In Figure~\ref{ona_cno} we summarize our results for C, N, O, and Na abundance
ratios, that have been plotted as a function of [(C+N+O)/Fe], [Fe/H] and
[La/Fe], for the first and second generation stars.
As well as O, carbon increases as a function of
the total CNO and [La/Fe], suggesting that C and O evolve in a similar
way (though their errors are correlated).

At odds with C and O, neither N nor Na show any evidence for
correlation with the total CNO abundance in first generation stars.
However, the minimum values for these two elements may increase slightly with both Fe
and La. 
N and Na display 
a much stronger trend with either the overall CNO, iron, and lanthanum
for second generation stars, i.e.\ N and Na are higher 
for stars with higher CNO/Fe/La.

All the observed abundance pattern suggests that C and O from one side,
and N and Na form the other, have undergone a similar processing in
the evolution of  the cluster.
In addition,  the first and second generation, selected on the basis
of their Na and O content, appear to show well defined individual
chemical trends along the overall observed range in metallicity.

Finally, the CNO increase among first generation stars is driven by
the rise of O (and partly also of C), rather than N. As C and O
increase with metallicity, more N could be produced by CNO cycling,
and so in O-poor/Na-rich second generation stars both [N/Fe] and
[Na/Fe] increase with metallicity. Figure~8 shows CNO vs. Fe
separately for first and second generation stars, demonstrating the
increase of CNO with Fe is common to both generations. The small
offset ($<\sim 0.1$ dex) between the two generations may be real (due
to part of CNO elements having been processed to heavier nuclei by
$p$-captures), but we cannot fully exclude a slight systematic
underestimate of N among second generation stars.

%
%

%
%
%
\begin{figure*}[ht!]
\centering
\epsscale{0.97}
\plotone{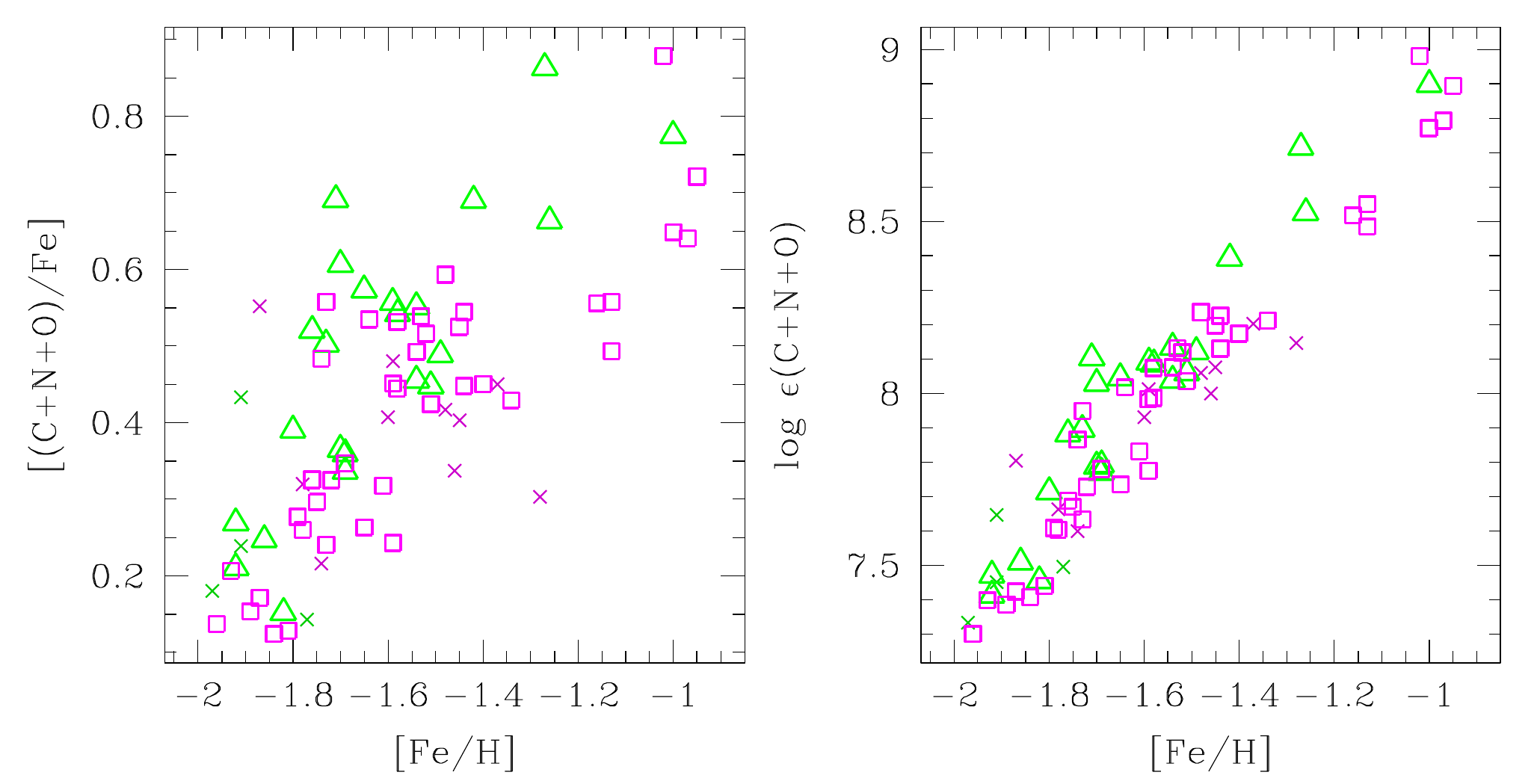}
\caption{CNO relative to Fe and log$\epsilon$(CNO) as a function of
 [Fe/H]. Symbols are as in Figure~\ref{ola}. Crosses mark stars for
 which we have only an upper limit for N or O.}
\label{cnoo}
\end{figure*}
%
\section{The formation and chemical enrichment of $\omega$~Cen}
\label{sec:scenarios} 
The interpretation of the photometric and
  spectroscopic evidences on the stellar populations of $\omega$~Cen
  in terms of its formation and evolution is posing formidable
  difficulties. In this Section we try to extract as much as possible
  from what the data themselves apparently demand.  

Although there is
  photometric evidence for at least six different stellar populations
  in $\omega$ Cen (e.g., Bellini et al. 2010), we have distinguished a
  {\it first} and a {\it second generation} as made of those stars
  that do not or do show evidence for additional $p$-capture
  processing at high temperature, respectively.  Thus, the {\it first
    generation} may represent a less complex case to interpret in
  terms of chemical evolution, with the {\it second generation}
  possibly arising from the AGB ejecta of the
former one, as generally entertained.

\subsection{Formation scenarios}
We first explore a scenario in which the progenitor system to
$\omega$~Cen has evolved as a single entity, in which the ISM was
chemically homogeneous at each time. Thus, one can envisage the first
generation to be the result of five or six successive bursts of star
formation, with stars in each successive episodes being progressively
enriched by nucleosynthesis products from core-collapse
supernovae from the previous bursts.
By
construction, the first generation is made of stars which formed out
of an ISM that was not yet polluted by $p$-capture products by AGB
stars. This requirement sets an upper limit of 30-40 Myr between the
first and the last burst, as later AGB stars would come at play.

Now, before the second generation starts to form most residual gas
must be ejected from the system, thus allowing second generation stars
being formed out of almost pure AGB ejecta. As emerging from all
previous figures, second generation stars span the full metallicity
range of $\omega$~Cen, which suggests that each episode of first
generation formation has produced its own specific contribution to the
second generation. Thus, after the five or six initial bursts of star
formation, and the ejection of the residual gas, a similar series of
bursts should have taken place out of gas from AGB ejecta being
accumulated at the bottom of the potential well, via a series of cooling flows.

Also for this second generation series of bursts one can set a time
limit to its duration, which comes from the necessity of avoiding the
contamination by C/O$>$1 materials from carbon star ejecta, which should
start some $2\times 10^8$ yrs after the very first burst. Therefore,
in this scenario all star formation episodes were confined to within
$\sim 2\times 10^8$ yrs.
Of course, such a scenario with its complex series of events may require yet
additional complications 
to explain the presence of 
the Na-O anticorrelation among stars within different individual star formation episode of the
second generation (i.e., within each [Fe/H] group).
Even more complications may be required if the
Na-O anticorrelation is due to partial mixing of AGB ejecta with
pristine material as often invoked (e.g., D'Ercole et al. 2011, and
references therein),
though the origin of the diluting material remains to be 
understood. Five or six successive dilutions with {\it pristine}
material would ask for an extremely contrived scenario. Alternatively,
common-envelope ejecta from intermediate-mass binary stars may offer a less
implausible option (Vanbeveren, Mennekens \& De Greve 2011)
but it remains to be seen if binaries can provide enough diluting material.

Moreover, as well known
the original first stellar generation must have been much more massive
than at present, if enough second generation stars formed from its AGB
ejecta (e.g., Bekki \& Norris 2006; Renzini 2008).
This led to the hypothesis of a dwarf galaxy
precursor of which $\omega$~Cen would be the stripped, remnant
nucleus (e.g., Bekki \& Freeman 2003; D'Ercole \etal 2010). 
In summary, in this scenario formation proceeds through a rapid series
of star formation episodes (bursts), separated by just few Myrs in
time, each followed by its own cooling flow made of AGB ejecta to
form the second generation stars in a series of secondary bursts. Tidal 
stripping will then remove most of the original stellar mass of the system,
leaving $\omega$~Cen as the bare nucleus of the original dwarf galaxy.

Alternatively, the several star
formation episodes,
 rather than being sequential in time may have been
separated in space, with each lump of matter having evolved
(quasi)independently from the others, before merging together to make
the cluster we observe today. In the original dwarf galaxy several
high-density sites of star formation were active, each forming a proto
globular cluster of first generation stars, with each of them later
feeding the formation of its own second generation stars via a cooling
flow of AGB-ejecta material.  In turn, the densest part of the
individual proto-clusters would merge together by dynamical friction
and coalesce, thus forming a massive nucleus, to become $\omega$ Cen
after all the rest of the dwarf galaxy is tidally stripped. 

Difficult to say whether the former, {\it monolithic} scenario is more
(or less) contrived than this {\it merger} one in which five or six
proto-clusters evolve separately before coalescing together. Yet, the
stellar population content of this most massive globular cluster in
the Galaxy is so complex that we suspect no simple, straightforward
model can be found for its formation. Eventually we must admit that we
are still far from understanding this most puzzling cluster.

\subsection{Chemical enrichment}

The puzzle becomes even more difficult to compose when considering the
specific chemical patterns exhibited by the stars in the cluster. Here
we limit the discussion to the CNO elements.

In Figure~8 [(C+N+O)/Fe] appears to increase with [Fe/H] among
first generation stars, a trend that is dominated by the increase of
[O/Fe] since oxygen is the most abundant of the three elements.
This
is also indicated by the tight correlation between [(C+N+O)/Fe] and
[O/Fe] seen in the right panel of Figure~6. 
The possible [O/Fe] increase with
[Fe/H] among first generation stars, is contrary to the
decreasing trend shown in all other environments.
In fact, [O/Fe] decreases in lockstep with [Fe/H]
in the Galactic thin disk, thick disk, bulge and halo (e.g.,
Bensby, Feltzing \& Lundstr{\"o}m 2004; Zoccali et al. 2006; Wheeler, Sneden \& Truran
1989), a trend that is generally interpreted as due to a prompt
enrichment of $\alpha$-element by core collapse supernovae, followed
by a slower iron enrichment by Type Ia supernovae (e.g., Matteucci \&
Greggio 1986).  Moreover, values of [O/Fe] in excess of 0.5 are common
among first generation stars, whereas they are extremely rare in any
other environment. Therefore, also the {\it normal} population in
$\omega$ Cen, the one not affected by proton captures, shows a unique
chemical evolution history, not paralleled in any other known
environment.

Thus, this classical scheme of chemical evolution may not apply to
$\omega$~Cen's first generation, as we have that oxygen keeps
increasing faster than iron, a clear evidence contrary to Type Ia
supernovae playing a role in its enrichment. 
If so, one has to resort
only on core collapse supernovae in trying to explain the [(C+N+O)/Fe]
trend with [Fe/H].  This is not an easy task.

The existence of a broad range of metallicities among the first
generation stars argues for enrichment and star formation having
proceeded together in the progenitor object of $\omega$~Cen, though
quite possibly as a series of successive {\it bursts}, as discussed in
the previous subsection and indicated by the distinct photometric
sequences present in this cluster (e.g., Bellini et al. 2010) paralleled
by a multimodal distribution of [Fe/H] (e.g., Marino et al. 2011a).

In the scenario in which star formation proceeds with a series of 
bursts, the first successive episode would have experienced an
enrichment due only to the most massive supernovae from the first
episode, and the following ones an enrichment due to a broader and
broader range of supernova masses. Thus a trend
in [O/Fe] with [Fe/H] could be explained {\it if} the ratio of
oxygen-to-iron yields from core collapse supernovae were a decreasing
function of stellar mass, i.e., if low mass supernovae were to make
more oxygen relative to iron, compared to high mass supernovae.

Theoretical yields of oxygen and iron as a function of the initial
mass of the supernova progenitor formally do not support this
possibility (see e.g. Woosley \& Weaver 1995; Limongi \& Chieffi 2006;
Kobayashi et al. 2006), but are sufficiently uncertain to leave this
option {\it a priori} relatively viable.
The main uncertainty comes from the iron yield as a function of stellar
mass being poorly constrained by supernova models, mainly due to the
difficulty in locating the {\it mass cut} between ejecta and compact
remnant that typically lies just within the iron layer of the
pre-supernova structure.  Nevertheless, this scheme would require
lower mass stars to dominate the oxygen production, whereas all
supernova models agree in predicting an increase of the oxygen yield
with stellar mass, at least up to $\sim 30~M_{\odot}$, and in most
models even beyond.

To explain the anomalous trend of [O/Fe] with [Fe/H] one may be
tempted to appeal to the last resort of desperate situations, and
invoke different IMFs for each successive episode of star formation, a
very arbitrary and contrived scenario indeed.
Moreover, note that this trend of increasing [O/Fe] with increase
[Fe/H] does not extend to the highest metallicity group in $\omega$
Cen, corresponding to the {\it anomalous} RGB-a and MS-a sequences in
the CMD (cf. old ref, Bellini et al. 2010), but here we have ascribed
this whole metallicity group to the $p$-capture processed second
generation.

In summary, even limiting ourselves to the {\it normal}, non
$p$-capture processed, first generation, we have to admit our failure
in providing an explanation for the observed [(C+N+O)/Fe] and [O/Fe]
trends with [O/Fe].  

\begin{figure*}[ht!]
\centering
\epsscale{.55}
\plotone{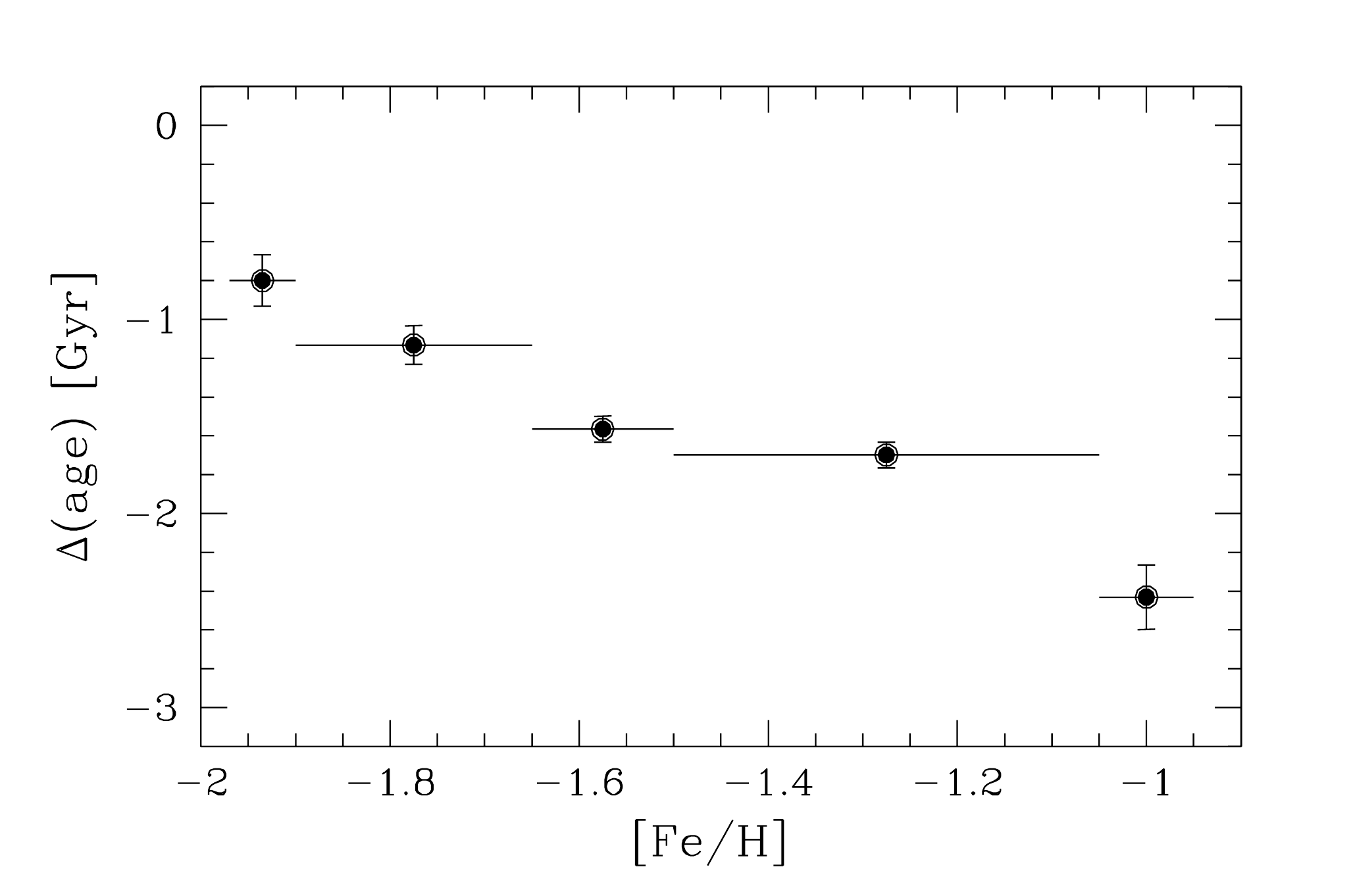}
\caption{Age difference for each $\omega$~Cen sub-population,
        between the estimates based on CNO-enhanced isochrones 
        and on the actual  C+N+O abundances, and the measurements 
        provided by not CNO-enhanced isochrones.} 
\label{agemet}
\end{figure*}
%

\section{CNO and age of $\omega$~Cen populations}
\label{sec:discussion}
The C+N+O abundance has a strong impact on the location of isochrones at 
the TO and SGB level, and therefore on age determinations.
In order to estimate the effects of the large variation of the overall CNO 
among $\omega$~Cen's stellar populations on the age measurement, we have 
calculated stellar models 
by including a variety of assumptions on the C+N+O abundance and
spanning a metallicity range suitable for the $\omega$~Cen stellar populations.     
The adopted physical inputs and 
numerical assumptions are the same as in Pietrinferni \etal (2009), 
which presents evolutionary computations accounting for a CNO enhancement 
of a factor of $\sim2$ relative to the \lq{standard}\rq\ $\alpha-$enhanced
mixture.  A fully consistent set of $\alpha$-enhanced isochrones with no 
CNO enhancement for various iron abundances (Pietrinferni \etal 2006), 
allows us to compare, in a fully homogeneous theoretical framework, 
CNO-enhanced and not-CNO-enhanced isochrones.

As expected, we find that, for a fixed TO and SGB brightness, the CNO-enhanced 
isochrones provide younger ages than isochrones corresponding to a canonical 
$\alpha$-enhanced mixture (Pietrinferni \etal 2006): as a rule-of-thumb 
we found the $\partial{\rm age}/\partial{\rm [CNO]}\sim-3.3$~${\rm Gyr~dex^{-1}}$, 
regardless of the iron content.  
By accounting for the observed [CNO/Fe] abundance 
in $\omega$~Cen, we have estimated the difference between the age one would
measure for each population 
by adopting appropriate CNO-enhanced isochrones and that obtained by using
\lq{normally}\rq\ $\alpha-$enhanced isochrones, i.e.\ neglecting the
measure CNO enhancement.
The overall effect is to decrease the age estimate for the higher
metallicity stars.
The age decrease ranges from $\sim$0.5~Gyr for stars with [Fe/H]$\sim
-$1.8  up to $\sim$2~Gyrs for the most metal-rich stars 
with [Fe/H]$\sim -$1.0 (Figure~\ref{agemet}),  as listed in
Table~3.

\begin{table*}[ht!]
\centering
\begin{tabular}{cccc}
\hline 
\hline
 [Fe/H]  & [(C+N+O)/Fe] & log$\epsilon$(C+N+O) & ${\rm \Delta{age}(Gyr)}$\\  
\hline
$-$1.97$<$[Fe/H]$<$$-$1.90 & 0.24$\pm$0.04 & 7.43$\pm$0.05 & $-$0.8$\pm$0.1   \\
$-$1.90$<$[Fe/H]$<$$-$1.65 & 0.34$\pm$0.03 & 7.69$\pm$0.04 & $-$1.1$\pm$0.1   \\
$-$1.65$<$[Fe/H]$<$$-$1.50 & 0.47$\pm$0.02 & 8.03$\pm$0.03 & $-$1.6$\pm$0.1   \\
$-$1.50$<$[Fe/H]$<$$-$1.05 & 0.48$\pm$0.04 & 8.24$\pm$0.05 & $-$1.6$\pm$0.1   \\
$-$1.05$<$[Fe/H]$<$$-$0.95 & 0.81$\pm$0.12 & 8.94$\pm$0.11 & $-$2.7$\pm$0.4   \\
\hline
\end{tabular}
\label{tab1}
\caption{CNO contents and estimated age differences for the
  $\omega$~Cen [Fe/H] groups. 
}
\end{table*}

This result confirms that the measurement of relative ages of stellar 
populations in GCs requires accurate measurements of the overall C+N+O 
abundance.  Indeed, it is well known that the double SGBs detected in NGC~1851, 
M22, 47~Tucanae (Milone \etal 2008, 2011, Anderson \etal
2009, Piotto 2009), can 
be interpreted in quite different ways, depending on the CNO content of the 
single populations.  In particular, the faint SGB can be associated either 
with a stellar population significantly older than the stars in the bright SGB 
(up to 1~Gyr), or to a second generation, almost coeval with 
the brighter SGB (age differences of 100-200~Myr), but with enhanced CNO 
(Cassisi \etal 2008, Ventura \etal 2009).  This scenario has received further 
support by recent findings in M22 by Marino \etal (2009, 2011b):  this GC hosts 
two stellar groups with different $s$-element and CNO abundances, and, in 
addition, the two groups of CNO-rich and CNO-poor stars have also different 
iron abundances, in close analogy with $\omega$~Cen. 
 
In NGC~1851, results on possible internal variations of C+N+O are contradicting
(Yong \etal 2009, Villanova \etal  2010).  However, theoretical models, 
properly accounting for photometric signatures of the chemical peculiarities
observed in the cluster sub-populations as provided by Sbordone \etal (2011),  
suggest that in NGC~1851 only a C+N+O-enhanced second generation can satisfy 
all the observational constraints given by CMDs in various photometric bands 
(Milone \etal 2008, Han \etal 2009). 

Many studies have attempted to reconstruct the evolutionary history of 
$\omega$~Cen, by determining relative ages among the different stellar 
sub-groups, but 
none has been able to account for variations in the overall C+N+O abundance.  
Considering the complexity of its SGB morphology,
and of the metallicity distribution, and different assumptions regarding He 
enhancement among $\omega$~Cen populations, the relative ages for the various 
stellar populations as measured in different studies are quite different, 
ranging from a small or null age dispersion, as proposed by 
Sollima \etal (2005), to an age spread of the order of 2-3~Gyr 
(Hilker \etal 2004, Stanford \etal 2006) to 4 or more Gyr 
(Hughes \& Wallerstein 2000, Hilker \& Richtler
2000, Villanova \etal 2007).

It is beyond the scope of this paper to evaluate in detail how C+N+O 
abundances change the age-dating obtained in the multiple cited works.
However, we can identify some general trends.  By using the same 
$\alpha$-enhanced isochrones adopted here,
Villanova \etal (2007) 
found a large age difference ($\sim$5 Gyrs) among the $\omega$~Cen 
populations.  Specifically they identify four groups of stars: 
(1) an old metal-poor group ([Fe/H]$\sim -$1.7), 
(2) an old population of metal-rich stars ([Fe/H]$\sim -$1.1), 
(3) a group of metal-intermediate young stars ([Fe/H]$\sim -$1.4), 
    about 1-2~Gyr younger than the old components, and 
(4) a young metal-poor group ([Fe/H]$\sim -$1.7) up to $\sim$5~Gyr 
younger than the old populations.  When accounting for the CNO overabundance 
and using appropriate CNO-enhanced 
isochrones, the age estimate of the metal-rich component in Villanova 
\etal should be decreased  by $\sim$2~Gyrs.  As a consequence, 
the metal-rich population is no longer coeval with the most metal-poor 
population, but is slightly younger.  Similarly, due to its enhanced 
CNO content, the intermediate metallicity  ([Fe/H]$\sim -$1.4) population 
would be younger by $\sim$1~Gyr relative to the estimates given by these 
authors.  On the other hand, the age difference between the two groups of 
old and young metal-poor stars should not change as they have the same 
iron and CNO content.

Hilker \etal (2004) and Stanford \etal (2006) suggest that $\omega$~Cen 
has experienced a  prolonged star-formation of $\sim$2-4~Gyrs but, at odds 
with Villanova \etal (2007), with the most metal-rich stars being
younger.  Similar conclusions have been reached either by 
Hughes \& Wallerstein (2000) and Hilker \& Richtler (2000), who estimated 
that the more metal-rich stars were younger than the metal-poor ones by 
$\geq$3~Gyrs, and 3-6~Gyrs respectively.  In the case of Hilker \etal (2004) 
and Stanford \etal (2006), we predict that CNO
effects on the relative-age estimate could increase the duration of the 
star-formation in $\omega$~Cen up to $\sim$4-6~Gyrs. 

Sollima \etal (2005) suggest that the global star-formation history of 
$\omega$~Cen should be confined to within $\sim$2~Gyrs, with the metal-poor 
population ([Fe/H]$\sim -$1.7) representing the first and the major
star-formation burst.  Our results suggest that the ages of the 
metal-intermediate and metal-rich stars given by these authors could be 
over-estimated by $\sim$0.6 and $\sim$1.5~Gyrs, with respect to the metal-poor 
population.  In any case, given their uncertainties of 2~Gyrs, these 
corrections do not change the main result of their work, e.g.\ the 
suggestion that $\omega$~Cen stellar populations formed in a short time period.

As mentioned above, the goal of this paper is not do a detailed analysis
of previous age determinations. 
As discussed before there are
  contradicting results in recent literature.
Our aim is simply to show that there is
significant variation in the C+N+O content among $\omega$~Cen's
populations that may easily change these datations.
In addition to that caveat, it is worth noting  that He enhancement is likely
present in the O-poor, metal-rich stars. This occurrence should be taken
into account when measuring the relative ages via isochrones fitting among
the various sub-populations, since He affects the SGB shape (but not the
luminosity level). 

\section{Conclusion\label{sec:conclusions}}
In this paper,  we have presented C, N, O abundances for 77 $\omega$~Cen 
RGB stars in the metallicity range $\sim -$2.0$<$[Fe/H]$< -$0.9. 
We have coupled our results with the ones of our previous work on
Na-O-La abundances presented in Marino \etal (2011a). 
From our analysis we found:
\begin{itemize}
\item {a correlation between the total CNO and the iron abundance, with the most 
metal-rich population being enhanced by $\sim$0.5 dex in [(C+N+O)/Fe] relative
to the most metal-poor one;}
\item{O and C grows with [Fe/H] and [La/Fe] in the O-rich/Na-poor stars.}
\item{stars selected on the basis of their position on the Na-O
    plane, show defined 
chemical patterns in their light elements 
C, N, O, and Na as a 
function CNO, Fe, and La,
that allow us to distinguish between a {\it first} and a {\it second
generation} of stars, each possibly resulting from a series of
separate bursts of star formation.}
\item{[La/Fe] correlates tightly with [Fe/H], following almost
precisely the same trend regardless of the O/Na abundances.}
\end{itemize}

In an attempt to make sense of this observational trends we have
explored two (speculative) scenarios for the formation and evolution
of this most puzzling object. In one option the system has evolved
{\it monolithically}, i.e., remaining chemically homogeneous at each
time.  Thus, within less than $\sim 30-40$ Myrs a series of bursts of
star formation each enriched in iron and $\alpha$ elements by
core-collapse supernovae from the previous burst(s) would have been
followed each by a secondary burst of star formation originated from
material ejected by AGB stars, heavily processed by $p$-captures.  The
formation of the whole stellar population inhabiting the cluster today
would have taken less than $\sim 2\times 10^8$ yrs.  Alternatively,
rather than sequential in time, such primary and secondary bursts of
star formation would have taken place separately in space, before
their products could merge at the bottom of the potential well.  In
both cases, secondary bursts would be fed via cooling flows made of
AGB ejects, leading to more centrally concentrated {\it second
generations}, which indeed appears to be so in this (and other)
clusters (Sollima \etal 2005, Bellini \etal 2009, Johnson \&
Pilachowski 2010).  We admit that both scenarios require a series of
very contrived and {\it ad hoc} assumptions, yet the extraordinary
complexity of $\omega$~Cen may not admit simple solutions.

The CNO abundance affects the determination of the relative ages of
cluster subpopulations via isochrone fitting of the turnoff SGB
region. In the light of our results, we have discussed this issue in
the case of $\omega$~Cen by comparing isochrones with standard and
with enhanced CNO, with the latter ones giving younger ages for the
same turnoff luminosity.  Although the determination of relative ages
of $\omega$~Cen sub-populations is beyond the aims of our study, we
argue that a trend in CNO/Fe can help reducing large age spread among
the various sub-populations, as found by some studies in the literature.

\begin{acknowledgements}
We thank the anonymous referee whose suggestions have significantly
improved this work.
APM, GP, SC and AA are founded by the Ministry of Science and
Technology of the Kingdom of Spain (grant AYA 2010-16717). APM and AA
are also founded by the Instituto de Astrofísica de Canarias (grant P3-94).
\end{acknowledgements}

\bibliographystyle{aa}

\end{document}